# Extending WS-Security to Implement Security Protocols for Web Services

Bela GENGE, Piroska HALLER

Department of Electrical Engineering, Faculty of Engineering, "Petru Maior" University of
Târgu Mureş, Mureş, e-mail: {bgenge, phaller}@engineering.upm.ro

**Abstract:** Web services use tokens provided by the WS-Security standard to implement security protocols. We propose several extensions to the WS-Security standard, including name types, key and random number extensions. The extensions are used to implement existing protocols such as ISO9798, Kerberos or BAN-Lowe. The advantages of using these implementations rather than the existing, binary ones, are inherited from the advantages of using Web service technologies, such as extensibility and end-to-end security across multiple environments that do not support a connection-based communication.

**Keywords:** Security protocols, Web services, WS-Security

## 1. Introduction

Security protocols are "communication protocols dedicated to achieving security goals" (C.J.F. Cremers and S. Mauw, 2005) [1] such as confidentiality, integrity or availability. Existing technologies such as the Security Assertions Markup Language [2] or WS-Security [3] provide a unifying solution for the authentication and authorization issues through the use of predefined protocols. By implementing these protocols, Web services authenticate users and provide authorized access to resources. However, in order to integrate new protocols, such as key-exchange or confidentiality protocols, we need to extend the WS-Security standard with new components.

In this paper we propose several extensions to the WS-Security standard including name types, key and random number extensions. The extensions were used to implement existing protocols such as ISO9798 [4], that makes use of the Diffie-Hellman [5] key exchange protocol with digital signatures, or the Kerberos V5 [6] symmetric key-based security protocol. The advantages of using these implementations rather than the existing, binary ones, are inherited from the advantages of using Web service technologies. From these we mention extensibility and end-to-end security across multiple environments that do not support a connection-based communication. In addition, by adding new tokens to the existing ones, message components can be further categorized and specialized, providing an increased security of these protocols because of the additional information that accompanies each component [9, 10].

The implementations were made according to the specifications of the SOAP [7] standard, which embodies the WS-Security components in its header. The execution timings revealed the possible use of these protocols in a wide variety of systems, ranging from e-commerce to multimedia streaming.

The paper is structured in four parts. After the introduction, section 2 illustrates the proposed extensions through the form of XML schemas. In section 3 we present our experimental results, clearly showing that the proposed extensions can be used to implement applications that require authentication, key exchange or confidentiality protocols. We end our paper with a conclusion and future work in section 4.

## 2. WS-Security extensions

WS-Security provides a set of tokens for implementing security properties such as authentication, integrity and non-repudiation [9, 11]. These properties are used by Web services to construct security protocols providing inter-domain authentication. These are predefined, static protocols that must be implemented by all communicating parties. In order to implement other authentication protocols or other types of security protocols, the tokens provided by WS-Security must be extended with several new ones.

We consulted a large number of security protocols from the SPORE [12] library and the library of protocols presented by John Clark [8]. Based on these, we identified four *basic sets* containing terms used by protocol participants to construct messages: $\mathsf{P}$, \textsf{N}; $\mathsf{N}$, denoting the set of nonces (i.e. "number once used"); $\mathsf{K}$, denoting the set of cryptographic keys and $\mathsf{M}$ denoting user-defined components.

The set of participant names $\mathsf{P}$ is further specialized with the following disjoint sets: $\mathsf{P}_{DN} \subseteq \mathsf{P}$, denoting the set of distinguished names; $\mathsf{P}_{UD} \subseteq \mathsf{P}$, denoting the set of user-domain names; $\mathsf{P}_{IP} \subseteq \mathsf{P}$, denoting the set of user-IP names; $P_D \subseteq P$, denoting the set of domain names; $P_U \subseteq P$, denoting the set of remaining user name types.

The set of nonces is also further specialized with two subsets: $N_R \subseteq N$, the set of random numbers and $N_T \subseteq N$, denoting the set of timestamps.

Based on the above-defined sets and subsets, in the remaining of this section we provide the XML representation of the terms corresponding to the implementation of each element. The WS-Security standard provides a single XML element for defining user names, through the form of *wsse:UsernameToken*. For example, in order to define a user name, the following syntax is required:

*<wsse:UsernameToken>Denumire utilizator</wsse:UsernameToken>*

Distinguished names are usually found in user certificates and they provide information related to the organization, country, domain and several other categories characterizing a user. In order to define user names in this format, we define the following XML schema:

*<complexType name="DistinguishedNameToken">*
    *<sequence>*
        *<element name="Organization" type="string"/>*
        *<element name="OrganizationalUnit" type="string"/>*
        *<element name="CommonName" type="string"/>*
        *<element name="Country" type="string"/>*
    *</sequence>*
*</complexType>*

User-domain names have the following structure: user@host.domain or user.host.domain. The schema for such a user name must include the user name and one or more host or domain names separated by dots. The resulting schema is the following:

*<complexType name="UserDomainNameToken">*
  *<sequence>*
    *<element name="UserName" type="string"></element>*
    *<element name="DomainName">*
      *<simpleType>*
        *<restriction base="string">*
          *<pattern value="(\w+\.\w+)+"></pattern>*
        *</restriction>*
      *</simpleType>*
    *</element>*
  *</sequence>*
*</complexType>*

IP addresses and identifying machine names must include support for both IPv4 and IPv6 address formats. The resulted XML schema makes use of regular expressions to describe the rules for constructing such names:

```
<complexType name="UserIPNameToken">
  <choice>
    <element name="IPV4">
     <simpleType>
      <restriction base="string">
            <pattern value="\d{1,3}\.\d{1,3}\.\d{1,3}\.\d{1,3}"/>
      </restriction>
     </simpleType>
    </element>
    <element name="IPv6">
     <simpleType>
      <restriction base="string">
         <pattern value="([0-9a-fA-F]{1,4}:){7}[0-9a-fA-F]{1,4}"/>
      </restriction>
     </simpleType>
    </element>
  </choice>
</complexType>
```

For names containing exclusive domain names, we use the following schema:

```
<simpleType name="DomainNameToken">
    <restriction base="string">
        <pattern value="(\w+\.\w+)+"></pattern>
    </restriction>
</simpleType>
```

Random numbers are transmitted as binary tokens, for which a security token is already provided by the WS-Standard. Transmitting timestamps is also possible by using existing tokens provided by WS-Security. However, in order to send and receive encrypted binary keys we use an XML schema that defines the key value and the encoding type used:

```
<complexType name="KeyToken">
  <sequence>
      <element name="KeyValue" type="string"/>
  </sequence>
  <attribute name="type">
      <simpleType>
       <restriction base="string">
          <enumeration value="base64Binary"/>
          <enumeration value="hexBinary"/>
```

```
        </restriction>
      </simpleType>
    </attribute>
  </complexType>
```

## 3. Experimental results

The proposed extensions were used to implement protocols with security properties ranging from authentication to key exchange and message confidentiality. The protocols were constructed from participants exchanging *terms*. Terms were constructed from the elements belonging to the basic sets provided in the previous section:

$$\mathsf{T} ::= . | \mathsf{P} | \mathsf{N} | \mathsf{K} | \mathsf{M} | (\mathsf{T},\mathsf{T}) | \{\mathsf{T}\}_{FuncName(\mathsf{T})}, \quad (1)$$

where *FuncName* defines the set of function names used to encrypt terms:

$$\begin{aligned}
NumeFunc ::= \ & sk \quad (symmetric encryption) \\
| \ & pk \quad (asymmetric\ encryption) \\
| \ & h \quad (hash\ encryption) \\
| \ & hmac \quad (keyed\ hash\ encryption)
\end{aligned} \quad (2)$$

By using the above definitions, protocol messages can be constructed as in the following examples:

- $\{A, B, N_a, K\}_{sk(K_{ab})}$, where $A, B \in \mathsf{P}$, $N_a \in \mathsf{N}$, $K \in \mathsf{K}$;

- $\{\{A, N_a\}_h\}_{pk(PK_a)}, A, N_a$, where $A \in \mathsf{P}$ şi $N_a \in \mathsf{N}$, $PK_a \in \mathsf{K}$.

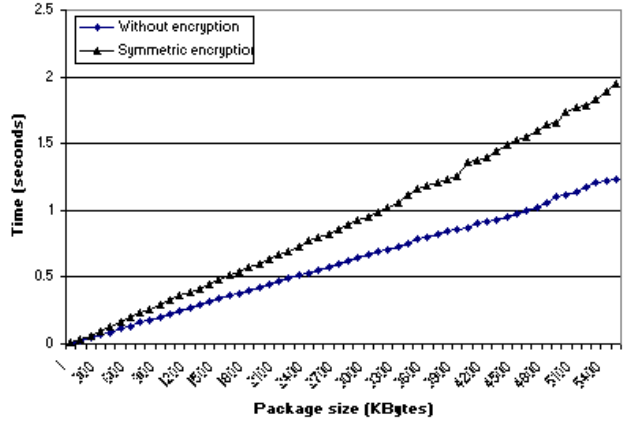

*Figure 1:*. Symmetric encryption versus no encryption.

The implementation of these messages replaces each component with its corresponding security token provided by the proposed extensions. The performance of the implementations is strongly dependent on the type of encryption function used in the process. For example, there is an obvious difference between an implementation that uses symmetric encryption and one that does not use encryption at all. This is the case illustrated in *Fig. 1*, where the encrypted message is $\{M\}_{sk(K_{ab})}$, with $M \in \mathsf{M}$. The figure illustrates the time required to construct, encrypt and send a message using the proposed tokens and the already existing ones.

In our experiments, messages were encoded in the SOAP [7] header, according to the WS-Security standard. Because of their size, as seen in *Fig. 1*, the XML structures influence the performance of the implemented protocols. This is also influenced by the type of encryption used, as shown in *Fig. 2*.

The illustrated values correspond to the execution time for constructed messages using symmetric and asymmetric cryptography. The symmetric encryption-based protocol is clearly much more performant than the asymmetric encryption-based protocol. This is why, the first protocol is usually used for data transfer, while the second one for encrypting small sized messages, usually in key exchange and authentication protocols.

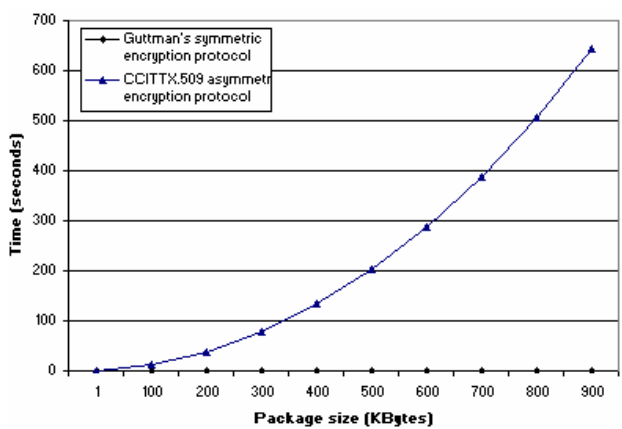

*Figure 2:*. Symmetric versus asymmetric encryption.

The experimental results given in Fig.1 and 2 show that the performance of the implemented protocols is not only influenced by the size of the encrypted messages, but also by the encryption algorithm type. We have implemented several other protocols, for which the execution timings are given in table 1. We identified several participants for each protocol. We measured the construction and the processing time of messages for each participant; the measured values were added together, resulting the total time.

We can see a clear difference between protocols that use symmetric algorithms (e.g. Lowe-BAN, Kerberos, Andrew RPC) and protocols that use asymmetric algorithms (e.g. ISO9798, CCITT X.509). For some protocols, the processing or construction timings are 0 because the sub-protocols we identified do not require operations. Based on these measurements, we can clearly state that using the proposed WS-Security extensions, we can implement key exchange, authentication and user-defined data exchange protocols. Implementing such protocols with existing WS-Security tokens is possible only for authentication protocols, for which the WS-Trust standard (using WS-Security) defines several predefined protocols.

*Table 1 :* Execution time of security protocols

| Participant role | Message construction (ms) | Message processing (ms) | Total participant (ms) | Total (ms) |
|---|---|---|---|---|
| Lowe-BAN Initiator | 11.81 | 3.68 | 15.49 | 19.97 |
| Lowe-BAN Respondent | 2.86 | 1.62 | 4.48 | |
| ISO9798 Initiator | 35.78 | 23.30 | 59.08 | 78.19 |
| ISO9798 Respondent | 6.87 | 12.24 | 19.11 | |
| Kerberos 1 Initiator | 0.83 | 0.00 | 0.83 | 27.69 |
| Kerberos 2 Initiator | 0.55 | 1.58 | 2.13 | |
| Kerberos 3 Initiator | 3.34 | 0.94 | 4.28 | |
| Kerberos 1 Respondent | 0.00 | 0.41 | 0.41 | |
| Kerberos 2 Respondent | 3.37 | 1.67 | 5.04 | |
| Kerberos 3 Respondent | 11.41 | 3.59 | 15 | |
| CCITT X.509 Initiator | 7.85 | 0.00 | 7.85 | 82.27 |
| CCITT X.509 Respondent | 0.00 | 74.42 | 74.42 | |
| Andrew RPC Initiator | 12.56 | 5.04 | 17.6 | 36.54 |
| Andrew RPC Respondent | 14.04 | 4.9 | 18.94 | |

## 4. Conclusions and future work

Existing tokens from the WS-Security standard provide the possibility for implementing a reduced set of security protocols. In order to enable the implementation of a wide range protocols, we proposed several token extensions for user name types and cryptographic keys.

The protocol implementations maintain their security properties by respecting the requirements given in the WS-Security standard. These requirements indicate the use of the SOAP header for transporting security tokens and the use of the SOAP body for other message components. The implementations we have developed show that protocol performance is influenced by the XML constructions and by cryptographic functions used in the process. Based on our experimental results, we can clearly state that the proposed extensions offer security for protocols used in various applications, such as multimedia or eCommerce.

In the future we intend to use the proposed extensions to implement several security protocols for multimedia applications and to prove that our implementations can be used to transfer audio and video messages without loss of quality.